# 2–16 GHz Multifrequency X-Cut Lithium Niobate NEMS Resonators on a Single Chip

Ryan Tetro, Luca Colombo, and Matteo Rinaldi
Institute for NanoSystems Innovation (NanoSI)
Northeastern University, Boston, MA, USA

*Abstract*—This work presents the design, fabrication, and testing of X-Cut Lithium Niobate (LN) acoustic nanoelectromechanical (NEMS) Laterally Vibrating Resonators (LVRs) and Degenerate LVRs (d-LVRs) operating in the $S_0$ (YZ30º) and $SH_0$ (YZ-10º) modes between 2 to 16 GHz range, monolithically fabricated on a single chip. The NEMS topology is optimized to extend the aforementioned fundamental modes in the C-, X-, and Ku-bands while preserving performance and mass-manufacturability. The devices present acoustic wavelengths ($\lambda$) varying between 1800 and 400 nm and are fabricated on a 100 nm ultra-thin LN film on high resistivity silicon with a 3-mask process. Experimental results highlighted quality factor at resonance ($Q_s$) and mechanical quality factors ($Q_m$) as high as 477 and 1750, respectively, and electromechanical coupling ($k_t^2$) as high as 32.7%. Large $k_t^2$ (> 10%) are recorded over a broad range of frequencies (2 – 6 GHz), while $Q_m$ exceeding 100 are measured up to 15 GHz. Further enhancement to performance and range of operation on the same chip can be achieved by decreasing $\lambda$, refining the fabrication process, and optimizing device topology. These additional steps can help pave the way for manufacturing high-performance resonators on a single chip covering the entire 1 – 25 GHz spectrum.

## I. INTRODUCTION

THE advent of 5G communication and Internet of Things (IoT) applications has led to an increased demand for high-performance piezoelectric resonators that can be scaled across the FR-1 (5 GHz), FR-2 (25 GHz), and proposed FR-3 (10 – 18 GHz) bands [1]. Key performance indicators (KPIs) crucial for RF communication and sensing include high-quality factor ($Q$) and robust electromechanical coupling ($k_t^2$). Furthermore, the increasing need for radio frequency front-end (RFFE) solutions would benefit from developing a monolithically integrated system that would respect cost and compactness constraints.

Current RFFE receivers implement surface acoustic wave (SAWs) and bulk acoustic wave resonators (BAWs) utilizing materials like scandium-doped aluminum nitride (ScAlN) [2] [3] or lithium niobate (LN) [4] [5]. Technologies proposed for X-band, Ku-band, and beyond [6] [7] exhibit a pronounced trade-off between $k_t^2$ and operating frequency, hindering multi-frequency filter bank synthesis on the same chip. Limitations in existing technologies, like ScAlN, prompt the exploration of alternatives. While ScAlN offers advantages [8] [9], attempts to boost its moderate $k_t^2$ leads to abnormally oriented grains (AOGs) in thin films, severely hindering the desired KPIs. Moreover, novel proposed methods, including overtones [10] or periodically polarized films (P3F) [6] [11] face challenges due to their intrinsic frequency-dependent behavior.

Scaling fundamental modes in nanoelectrical mechanical systems (NEMS) to higher frequencies is a straightforward solution but presents some formidable challenges. Commercially available devices such as Film Bulk Acoustic Resonators (FBARs) or recently developed Cross-Sectional Lamé Mode Resonators (CLMRs) [12] either showcase the $k_t^2$ required for FR-2 and FR-3 operation, but lack lithographic tunability for broad filter bank synthesis or do not possess KPIs compatible with technological requirements. Additionally, increasing the frequency of operation requires tight fabrication constraints, impacting device performance and yield.

This work addresses these challenges by exploring Lamb Wave NEMS devices, specifically $S_0$ and $SH_0$ modes in ultra-thin film X-Cut Lithium Niobate Laterally Vibrating Resonators (LVRs). Unlike other technologies, suspended Lamb Wave devices support lithographically defined multi-frequency on the same chip with limited Figure of Merit (FoM) degradation across a broad spectrum of frequencies [15]. Leveraging these capabilities, in combination with the intrinsic large LN piezoelectric coupling, the resonators showcased here exhibit high $k_t^2$ over an extensive frequency range within a single chip.

Furthermore, a degenerate LVR (d-LVR) configuration is introduced to alleviate the lithographic constraints and increase the operational frequency.

The following sections delve into the physics, fabrication process, and performance of $S_0$ and $SH_0$ mode resonators, utilizing COMSOL® Finite Element Analysis (FEA) models to illustrate the relationship between resonator design and attainable $k_t^2$. The admittance and performance of each resonator are reported, accompanied by insights into potential pathways for further improvement in key performance metrics.

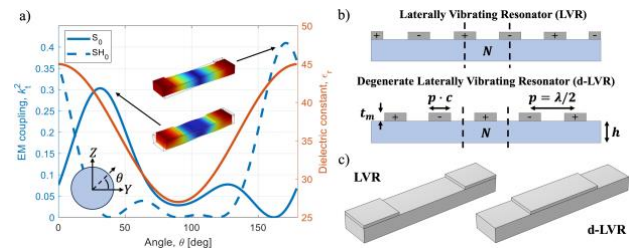

Fig. 1. a) Simulated electromechanical coupling for the $S_0$ and $SH_0$ modes (whose displacement at resonance is captured in the insets), and the relative dielectric constant as a function of the direction of propagation ($\vartheta$) according to [13] [14]. $\vartheta$ is defined as the angle between the Y and Z axes on X-cut LN; b) Electrode configuration for the LVR and d-LVR, constituted by a number of elements ($N$) of 5; and c) COMSOL® 2.5D model unit cell with a metallization ratio (or coverage, $c$) of 50%. Free and continuous boundaries are set in the lateral and transverse directions, respectively.

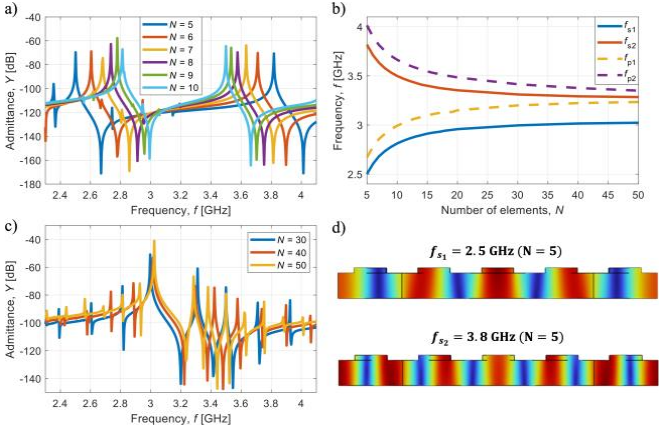

Fig. 2. a) COMSOL® FEA $S_0$ d-LVRs simulated admittance responses with number of arrayed elements ($N$) ranging between 5 and 10, $\lambda$ = 2 $\mu$m, $h$ = 100 nm, $t_m$ = 20 nm, and $c$ = 50%. d-LVRs show a split peak due to non-optimal boundary conditions [18] [19]; b) Simulated resonance and anti-resonance frequencies shift as a function of $N$. Both $f_s$ and $f_p$ show a convergent behavior as $N$ increases; c) For large $N$ (> 50), the boundary conditions do not impart a significant shift in the excited modes from the designed $\lambda$. However, multiple peaks appear due to the d-LVR electrode configuration. d) FEA simulated displacement at resonance for $f_{s1}$ and $f_{s2}$ and $N$ = 5. The boundary conditions force the excitation of modes whose wavenumber ($k_x$) does not match the electrodes' periodicity. In this case, two modes with 4 and 6 displacement nodes are excited by the d-LVR configuration. The displacement fields are scaled in z by 4x for clarity.

## II. METHODS

This study exploits the $S_0$ mode in laterally excited X-cut YZ30° LN and the $SH_0$ mode in laterally excited X-cut YZ-10° LN for synthesizing acoustic nanoresonators suitable for 5G and beyond-5G applications. The resonator structure comprises a thin film LN plate featuring top metal electrode Inter Digitated Transducers (IDTs), bus, and anchor [16]. Acoustic waves are generated by applying an RF voltage to the IDTs between device terminals, inducing a time-varying stress in the piezoelectric medium. The desired resonance is chosen by spacing the IDTs of opposing terminals by half the wavelength ($\lambda/2$), enabling design flexibility for multiple resonators on the same substrate, as long as the desired mode maintains a real wave number ($k_x$) within the dispersion range and sufficient electromechanical coupling ($k_t^2$) [17].

### A. LVR and d-LVR Topologies

The $S_0$ and $SH_0$ LVRs exploit the $e_{11}$ and $e_{15}$ piezoelectric coefficients, respectively, placing the point of maximum stress and strain between IDTs of opposite terminals. To satisfy the zero-stress, maximum displacement boundary conditions in the lateral direction, it is necessary to include a half-width IDT at the edge of the resonator plate, as shown in Fig. 1b [18]. However, lithographic constraints arise due to this half-width IDT, particularly at higher frequencies. To address this, a modified top electrode IDT configuration is implemented, specifically a degenerate LVR (d-LVR) topology [19]. The d-LVR inherits the top electrode configuration of cross-field excited Lamb wave resonators (LWRs), as displayed in Fig. 1b [20]. To satisfy the zero-stress, maximum displacement boundary condition in this class of devices, the edge of the resonator plate must be spaced $\lambda/4$ from the center of the adjacent IDT [21].

### B. d-LVR FEA Modeling

Past investigations of degenerate LVRs [18] [19] have showcased split resonance peaks with reduced coupling [22]. This phenomenon is ultimately induced by their non-ideal IDT configuration for in-line excitation, which does not allow the correct spatial frequency sampling of a standing wave possessing the number of displacement nodes equal to the number of electrodes' pairs ($N$). The $S_0$ FEA simulations reported in Fig. 2a demonstrate these findings. However, for larger $N$, the resonance and antiresonance frequencies of the split peaks ($f_s$ and $f_p$, respectively) asymptotically converge towards the frequencies of the non-degenerate IDT configuration (Fig. 2b), at the expenses of inducing an increased number of peaks close to the desired mode (Fig. 2b). This is caused by the increased lateral plate dimension compared to the IDT spacing, allowing for a greater number of modes that can exhibit non-null coupling with the degenerate electrode configuration.

### C. Fabrication

LVR and d-LVR NEMS resonators are fabricated on 100 nm ultra-thin film X-cut LN on high-resistivity silicon. To maximize the $k_t^2$, the $S_0$ resonators are oriented at YZ30° [13] and the $SH_0$ resonators are oriented at YZ-10° [14]. The fabrication process involves three levels of lithography to define the resonator plate, the Aluminum Silicon Copper (AlSiCu) IDTs, and the Gold (Au) contact pads (Fig. 3). The acoustic cavity is defined using direct-write lithography, followed by a high-density argon-based reactive ion etch. Nanometer-scale AlSiCu IDTs are patterned through electron beam lithography and liftoff. The Au contact pads are aligned to the AlSiCu IDTs and patterned through direct-write lithography and liftoff. Finally, the resonator plate is released from the silicon substrate via an isotropic $XeF_2$ etch.

## III. MEASUREMENTS

330 $S_0$ and 330 $SH_0$ NEMS resonators operating between 2 GHz and 16 GHz were successfully fabricated on ultra-thin film LN. Of the 330 resonators for each respective mode,

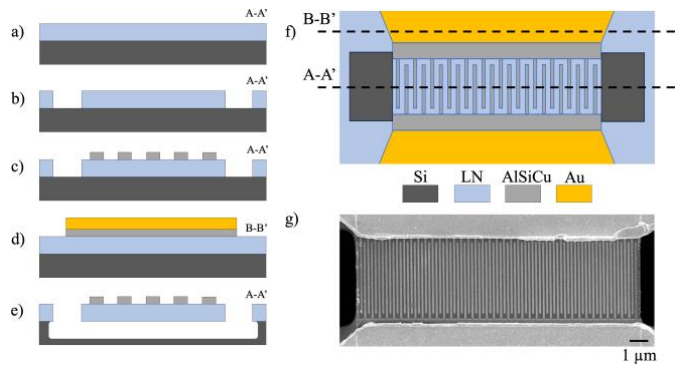

Fig. 3. Nanoresonator fabrication process a) starting with 100 nm thin film LN on Si, provided by NGK; b) the LN is etched to define the resonant cavity; c) the IDTs and d) contact pads are patterned through two separate liftoff processes via e-beam and direct lithography, respectively; and e) the resonator plate is released with a $XeF_2$ etch. This simple process can be scaled to mass manufacture devices with high $Q_s$ and $k_t^2$ with the use of a deep UV lithography stepper; f) Schematic top view of a d-LVR; and g) An SEM image of a $S_0$ resonator operating around 16 GHz (YZ30°). The same structure, manufactured with a direction of propagation YZ-10°, would maximize $k_t^2$ for the $SH_0$ around 9 GHz.

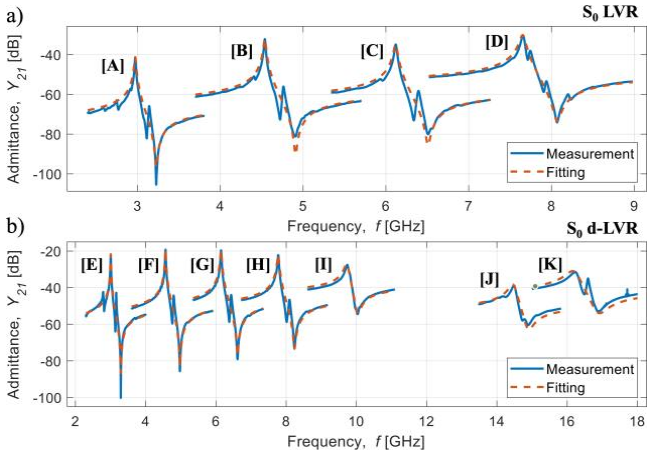

Fig. 4. a) $S_0$ LVR admittance plots and single-mode Butterworth-Van Dyke (BVD) fitting; and b) d-LVR admittance plots and single-mode BVD fitting. The BVD and multi-MBVD performance results are reported in Tables I and II.

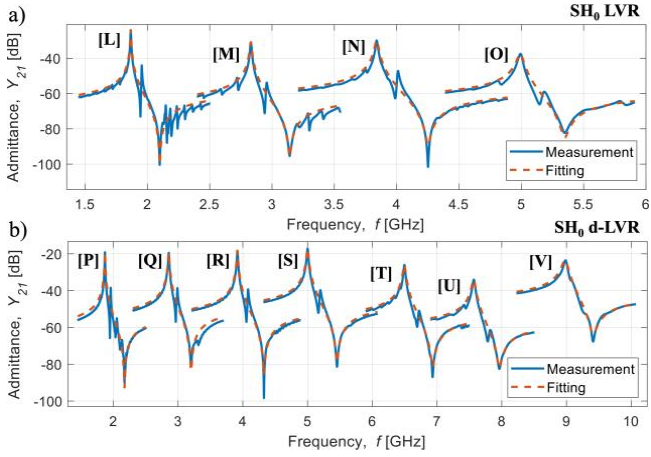

Fig. 5. a) $SH_0$ LVR admittance plots and single-mode Butterworth-Van Dyke (BVD) fitting; and b) d-LVR admittance plots and single-mode BVD fitting. The BVD and multi-MBVD performance results are reported in Tables III and IV.

TABLE I
$S_0$ LVRs

|   | $\lambda$ [nm] | $f_s$ [GHz] | $Q_s$ | $Q_p$ | $Q_m$ | $k_t^2$ | $C_0$ [fF] | FoM |
|---|---|---|---|---|---|---|---|---|
| A | 1800 | 2.99 | 261 | 549 | 997 | 18.4% | 17.6 | 48 |
| B | 1200 | 4.54 | 288 | 84 | 316 | 15.9% | 28.2 | 46 |
| C | 900 | 6.12 | 234 | 82 | 247 | 12.7% | 23.8 | 30 |
| D | 720 | 7.65 | 187 | 197 | 230 | 8.5% | 50.7 | 13 |

TABLE II
$S_0$ D-LVRs

|   | $\lambda$ [nm] | $f_s$ [GHz] | $Q_s$ | $Q_p$ | $Q_m$ | $k_t^2$ | $C_0$ [fF] | FoM |
|---|---|---|---|---|---|---|---|---|
| E | 1800 | 3.00 | 316 | 1383 | 1318 | 19.9% | 93.6 | 63 |
| F | 1200 | 4.56 | 321 | 393 | 505 | 18.8% | 89.8 | 60 |
| G | 900 | 6.15 | 277 | 297 | 335 | 16.0% | 93.1 | 44 |
| H | 720 | 7.77 | 259 | 245 | 230 | 11.7% | 79.7 | 30 |
| I | 560 | 9.74 | 101 | 100 | 105 | 6.9% | 148 | 7 |
| J | 480 | 14.47 | 105 | 56 | 121 | 4.0% | 31.4 | 4 |
| K | 400 | 16.21 | 55 | 56 | 71 | 5.8% | 71.7 | 3 |

TABLE III
$SH_0$ LVRs

|   | $\lambda$ [nm] | $f_s$ [GHz] | $Q_s$ | $Q_p$ | $Q_m$ | $k_t^2$ | $C_0$ [fF] | FoM |
|---|---|---|---|---|---|---|---|---|
| L | 1800 | 1.87 | 477 | 588 | 1143 | 29.7% | 51.4 | 142 |
| M | 1200 | 2.83 | 301 | 228 | 548 | 24.4% | 28.2 | 73 |
| N | 900 | 3.84 | 242 | 750 | 490 | 19.5% | 35.5 | 58 |
| O | 720 | 4.99 | 158 | 107 | 196 | 13.7% | 24.0 | 22 |

TABLE IV
$SH_0$ D-LVRs

|   | $\lambda$ [nm] | $f_s$ [GHz] | $Q_s$ | $Q_p$ | $Q_m$ | $k_t^2$ | $C_0$ [fF] | FoM |
|---|---|---|---|---|---|---|---|---|
| P | 1800 | 1.87 | 400 | 481 | 1750 | 32.7% | 99.2 | 137 |
| Q | 1200 | 2.86 | 262 | 212 | 1368 | 29.3% | 111.7 | 77 |
| R | 900 | 3.92 | 332 | 1374 | 1219 | 23.7% | 86.1 | 78 |
| S | 720 | 5.00 | 277 | 335 | 904 | 20.1% | 103.6 | 56 |
| T | 560 | 6.50 | 299 | 488 | 694 | 14.0% | 41.7 | 42 |
| U | 480 | 7.57 | 239 | 199 | 285 | 11.1% | 22.5 | 27 |
| V | 400 | 8.98 | 163 | 244 | 244 | 9.2% | 106.1 | 15 |

120 were LVRs with acoustic wavelengths ranging between $\lambda$ = 1800 nm and $\lambda$ = 720 nm, and 210 were d-LVRs with acoustic wavelengths ranging between $\lambda$ = 1800 nm and $\lambda$ = 400 nm. Numerous variations of the LVR and d-LVR typologies were designed with different aperture lengths ($L_e$) and number of IDT finger pairs ($N_p$) [16] to identify a design that maximizes both $Q$ and $k_t^2$ at each frequency. These metal traces electrically drive the resonator through the probing contact pads connected to the resonator anchor, bus, and IDTs. An SEM image of a fabricated resonator is reported in Fig. 3 with dimensions $L_e$ = 10 $\lambda$ and $N_p$ = 80.

Measurements are conducted using a VNA probe station equipped with GSG-150 probes, calibrated with a SOLT calibration for the desired frequency range to evaluate the resonators' admittance frequency response. Each resonator is individually probed in a 2-port configuration to extract the through admittance ($Y_{21}$) response around the designed resonant frequency. The data for each resonator is saved as $S$-parameters and later converted to $Y$-parameters via software to be individually analyzed for performance.

The admittance curves for all the best-performing LVRs an d-LVRs at each frequency were fitted using a Butterworth-Van Dyke (BVD) equivalent circuit model (Fig. 4 and Fig. 5) to extract the resonance $Q$ ($Q_s$), the antiresonance $Q$ ($Q_p$), and the static capacitance ($C_0$). Additionally, a multi-Modified Butterworth-Van Dyke (MBVD) fitting was utilized to fit the spurious modes and extract the mechanical $Q$ ($Q_m$) and the intrinsic $k_t^2$.

## IV. RESULTS AND DISCUSSION

The results of the BVD and multi-MBVD analysis, shown in Tables I, II, III and IV, demonstrate a maximum Figure of Merit (FoM) = 142, $Q_s$ = 477, $k_t^2$ = 32.7%, and $Q_m$ = 1750 at 1.87 GHz. For the higher frequency resonators, both $S_0$ and $SH_0$ modes maintain $k_t^2$ > 10% across the 2-8 GHz range, with $Q$ > 100 sustained up to 14 GHz. Though the LVRs are expected to outperform the d-LVRs, fabrication-related issues, particularly in resolving half-width edge electrodes, likely contribute to the degradation of LVR performance. Conversely, the d-LVRs showcase impressive results due to the large number of IDT pairs and the effective suppression of unwanted modes likely resulting from the lack of precise acoustic cavity definition during fabrication. d-LVRs operated with comparable performance in the 2-8 GHz range while showcasing a drastic reduction in $Q$ and $k_t^2$ once approaching the Ku-band (10 GHz). This phenomenon is likely tied to the extremely narrow patterned features, which can be addressed by design and fabrication process improvements.

## V. CONCLUSIONS

Laterally Vibrating Resonators (LVRs) and degenerate Laterally Vibrating Resonators (d-LVRs) operating in the $S_0$ and $SH_0$ modes between 2 to 16 GHz are designed, fabricated, and tested on a single chip with a 3-mask process. Experimental results highlighted large electromechanical couplings over a broad frequency range and good quality factors up to 14 GHz. Upon further development and optimization, the authors believe that the described technology, supporting large $k_t^2$ over the whole C-, X-, and Ku- bands, is a promising candidate to tackle the hardware challenges posed by novel communication paradigms and possibly extend to the mmWave range.

## ACKNOWLEDGMENTS

This work was supported by funding from the U.S. Army Combat Capabilities Development Command Research Laboratory (ARL Devcom) and Rogers Corporation. The authors would also like to thank Northeastern University Kostas Cleanroom and Harvard CNS staff.